\documentclass[fleqn,usenatbib]{mnras}

\usepackage[T1]{fontenc}

\DeclareRobustCommand{\VAN}[3]{#2}
\let\VANthebibliography\thebibliography
\def\thebibliography{\DeclareRobustCommand{\VAN}[3]{##3}\VANthebibliography}

\usepackage{graphicx}         
\usepackage{amssymb, amsmath} 
\usepackage{subfig}           

\title{Time delay estimation in unresolved lensed quasars}

\author[L.~Biggio et al.]{
  L.~Biggio,$^{a}$
  A.~Domi,$^{b}$
  S.~Tosi,$^{c}$
  G.~Vernardos,$^{d}$
  D.~Ricci,$^{e}$
  L.~Paganin,$^{c}$
  G.~Bracco$^{c}$
  \\
  $^{a}$ Eidgen\"ossische Technische Hochschule Z\"urich, R\"amistrasse 101, CH-8092 Z\"urich, Switzerland \\
  $^{b}$ University of Amsterdam, Science Park 105, 1098 XG Amsterdam, the Netherlands \\
  $^{c}$ Università degli Studi di Genova and Istituto Nazionale di Fisica Nucleare (INFN) \\- Sezione di Genova, via Dodecaneso 33, 16146, Genoa, Italy \\
  $^{d}$ Institute of Physics, Laboratory of Astrophysics, Ecole Polytechnique F\'{e}d\'{e}rale de Lausanne (EPFL), \\ Observatoire de Sauverny, 1290 Versoix, Switzerland \\
  $^{e}$ Istituto Nazionale di Astrofisica (INAF), Osservatorio di Padova - Vicolo dell'Osservatorio, 5, 35122 Padova, Italy \\
}

\begin{document}
\label{firstpage}
\pagerange{\pageref{firstpage}--\pageref{lastpage}}
\maketitle

\begin{abstract}
Time-delay cosmography can be used to infer the Hubble parameter $H_0$ by measuring the relative time delays between multiple images of gravitationally-lensed quasars. 
A few of such systems have already been used to measure $H_0$: their time delays were determined from the multiple images light curves obtained by regular, years long, monitoring campaigns. Such campaigns can hardly be performed by any telescope: many facilities are often over-subscribed with a large amount of observational requests to fulfill. While the ideal systems for time-delay measurements are lensed quasars whose images are well resolved by the instruments, several lensed quasars have a small angular separation between the multiple images, and would appear as a single, unresolved, image to a large number of telescopes featuring poor angular resolutions or located in not privileged geographical locations. Methods allowing to infer the time delay also from unresolved light curves would boost the potential of such telescopes and greatly increase the available statistics for $H_0$ measurements. This work presents a study of unresolved lensed quasar systems to estimate the time delay using a deep learning-based approach that exploits the capabilities of one-dimensional convolutional neural networks. Experiments on state-of-the-art simulations of unresolved light curves show the potential of the proposed method and pave the way for future applications in time-delay cosmography.
\end{abstract}


\section{Introduction}

\begin{figure}
  \centering
  \includegraphics[width=\columnwidth]{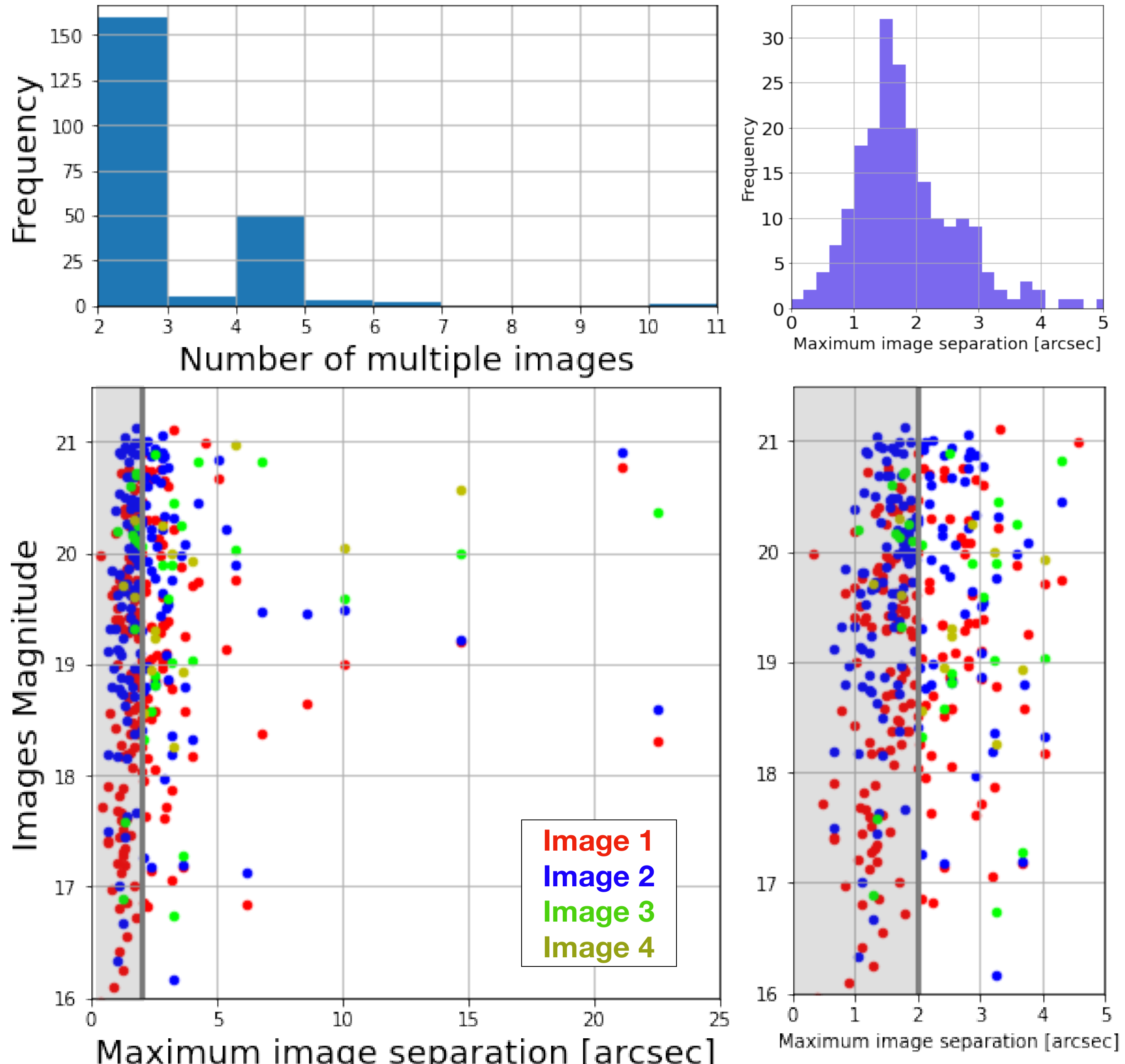}
  \caption[cm]{Number of images and image separation for the population of known gravitationally-lensed quasars. Top-left: distribution of known GLQs as a function of
    the number of multiple images. Top-right: distribution of known
    GLQs as a function of the maximum image separation. Bottom: Magnitude of the multiple images versus the maximum
    image separation for lensed systems with up to 4 multiple images (left); a zoom in the region 0-5 arcsec is shown on the right. The different colors of the dots identify each of the multiple images, from 1 to 4. The grey region contains 70$\%$ of the total GLQ
    sample.}
  \label{fig:qsosepmagboth}
\end{figure}

The Hubble parameter $H_0$, quantifying the
current expansion rate of the universe, is a major component of cosmological models, which can then be tested by its determination. To date, measurements of $H_0$ from different observations have led to a tension on its estimated value. In particular, early universe observations of the CMB anisotropies \citep[e.g., from the Planck satellite,][]{planck} have measured $H_0 = 67.4 \pm 0.5$ km s$^{-1}$ Mpc$^{-1}$, whereas late universe probes such as the distance ladder \citep{cepheid_H0} give $H_0 = 74.03 \pm 1.42$ km s$^{-1}$ Mpc$^{-1}$, resulting in a tension of about $4.4\sigma$ \citep{DiValentino, Beenakker, Verde}.
As firstly pointed out by \cite{Refsdal1964}, an additional method to determine $H_0$ is time-delay cosmography, which exploits that fact that the time delay ($\Delta T$) between multiple images of gravitationally lensed quasars (GLQs) is directly related to the Hubble parameter.
The most relevant results obtained via time-delay cosmography come from the H0LiCOW collaboration \citep{H0licow_XIII}, who has found $H_0 = 73.3_{-1.8}^{+1.7}$ km s$^{-1}$
Mpc$^{-1}$ from a sample of six GLQs monitored by the COSMOGRAIL
project \citep{Cosmograil2020}. This result, combined with the other
late universe observations \citep{cepheid_H0}, enhances the $H_0$
tension up to $5.3\sigma$. However, a more recent analysis of 40
strong gravitational lenses, from TDCOSMO+SLACS \citep{tdcosmo}, has
found $H_0 = 67.4_{-3.2}^{+4.1}$ km s$^{-1}$ Mpc$^{-1}$, relaxing the
tension and demonstrating the importance of a better understanding of
the mass density profiles of the lenses. In this context, further
studies including more objects are needed for a more precise estimation of the $H_0$
parameter \citep{tdcosmoV}.
The fractional error on $H_0$ for an ensemble of $N$
GLQs is related to the uncertainties in the time-delay measurement,
$\sigma_{\Delta T}$, line-of-sight convergence, $\sigma_{\rm los}$,
and lens surface density, $\sigma_{[k]}$, as \citep{Tie_2017}:
\begin{equation}
\frac{\sigma_H^2}{H_0^2} \sim \frac{\sigma_{\Delta T}^2/\Delta T^2 + \sigma_{\rm los}^2 + \sigma_{[k]}^2}{N},
\end{equation}
where the first two terms are dominated by random uncertainties and
their contributions scale as $N^{-1/2}$. There are therefore two ways of reducing the uncertainty on $H_0$: 1) reduce the contribution of random uncertainties, 2) increase the
size $N$ of the analysed GLQ sample.
The main contribution on random uncertainties is given by the microlensing effect \citep{Tie_2017}: massive objects such as giant stars, black holes, etc, present in the lensing system, can partially absorb, deflect or magnificate the light coming from the source. This results in changes in the light curves which can mistakenly be exploited to estimate $\Delta t$. 
With respect to the size $N$, to date, a sample of about $220$ GLQs is
available\footnote{\url{https://research.ast.cam.ac.uk/lensedquasars/index.html}}, however, only a very small subset with well separated multiple images has been used to measure $H_0$. Indeed, larger-separation systems benefit of better resolved space-based data, which in turn allow for better constraints on the mass models; moreover, it is easier to monitor brighter systems and to obtain their time delays. Therefore, it is easier and safer to extract information from such systems, and consequently reduce the uncertainty on $H_0$. Figure
\ref{fig:qsosepmagboth} (bottom) shows the magnitude of the multiple
images versus the maximum image separation for the known GLQs. Systems
falling in the grey region, which represents 70$\%$ of the total
sample, have a maximum image separation below 2 arcsec. The image
separation peaks indeed at around 1 arcsec \citep{OguriMarshall, Collett}, making the smaller and harder to observe systems the most numerous GLQs in future surveys.

%

The ideal instruments to perform lensed quasar monitoring have high sensitivity, an optimal geographical location (where the effects of atmospheric turbulence are less prominent), and a high angular resolution optimized with the usage of state-of-the-art adaptive optics systems. However, because of the time
scales of the intrinsic variations of the sources, which can be of the
order of years, such observation campaigns should last several
observing seasons \citep{Cosmograil2020}. Consequently, due to the
amount of observational requests that the best performing telescopes have to fulfill,
they can hardly be employed for such monitoring purposes. On the other
hand, small/medium sized telescopes ($\approx1$-$2$m or smaller) can often guarantee a better availability of
observational time for this purpose \citep{Borgeest1996}. Unfortunately, their already reduced sensitivity can be further worsened by their often less privileged geographical
positions, in terms of clear nights and atmospheric seeing, which can
reach up to 3 arcsec \citep{karttunen2016fundamental}.  While a few lensed quasars can be fully resolved by such facilities, and indeed time delay curves have been provided by them (for example the 1.2 m Euler Swiss telescope at ESO La Silla), still the majority
of the already known GLQs, together with future discoveries, will
mainly appear as a single image for these telescopes.

The identification of more strongly lensed quasars from unresolved light curves can clearly boost the outcome of upcoming surveys but it represents a challenging problem because of the limited angular resolution of wide surveys: proposals have been made to identify lensed systems even from not fully resolved light curves (see for example \citep{ShuY}). Light curves from resolved sources are then  analysed using point estimators to derive time delays \citep{CosmograilXI}; a recent proposal was advanced to deal with the unresolved cases, based on minimizing fluctuations in the reconstructed light curves \citep{Bag}. 

This work proposes a novel approach to estimate the time delay in unresolved GLQ light curves based on deep learning algorithms. Deep Learning (DL) is an emerging field of Machine Learning that has reached state-of-the-art performance in several applications. Cosmology and astrophysics will also benefit from the application of deep learning techniques, in particular in light of the need for more efficient data analysis tools and the unprecedented amount of information expected from the launch of several upcoming surveys, such as the Vera Rubin Observatory \citep{LSST}.

For simplicity, the usage of DL on GLQ light curves in this work only focuses on unresolved image pairs, which can come either from doubly- or  quadruply-lensed GLQs. This choice
is further motivated by the fact that about 85 per cent of the already
known systems are doubles \citep{OguriMarshall, Collett}, as shown in
Fig. \ref{fig:qsosepmagboth} (top-left).

The paper is structured as follows: Sec. \ref{sec:idea_and_method
  description} describes the deep learning-based method used for
evaluating the time delay between unresolved multiple images, Sec. \ref{sec:MC}
describes generating the simulated light curves needed for training the DL
algorithm, and Sec. \ref{sec:test_approach} shows the results of the
proposed method on a test dataset.


\section{Time delay estimation with deep learning}
\label{sec:idea_and_method description}

The method we adopt here exploits the ability of modern Convolutional Neural Networks (CNNs) \citep{cnn1} to extract informative features directly from raw data; they work in an end-to-end fashion given a supervised-learning task of interest that utilizes a pre-trained model. In our case, this task is the estimation of the time-delay between two unresolved quasar light curves.


The approach is motivated by the surprisingly good performance of
Machine Learning and, in particular, Deep Learning methods in a wide range of engineering fields, including astrophysics and cosmology: in addition to automated tasks on the large datasets of wide survey experiments (see for example \citep{Cabrera, Kimura, George, Schawinski, Sedaghat, Shallue}), deep learning is also proposed to analyse time series (for example \cite{REIMERS2020251, Wei}).
  
Most of these techniques are based on the supervised learning
paradigm, i.e. when labelled data are available and the algorithm can rely on explicit supervision signals. In the case of most deep
learning algorithms, the extent of such supervision is often
significant, meaning that large labelled datasets are needed for
effective learning. This scenario often results in excellent
performances when labelled data are abundant and their collection is
easy and not expensive. However, these conditions are not always 
satisfied and, in absence of aforementioned labelled datasets, one
must resort to either unsupervised or self-supervised learning
strategies, for which only unlabelled data are used, or to synthetic
data generation in order to produce the desired labelled datasets in
such a way that the artificial data resemble the real ones as much as
possible.

\begin{figure}
  \centering
  \includegraphics[width=\columnwidth]{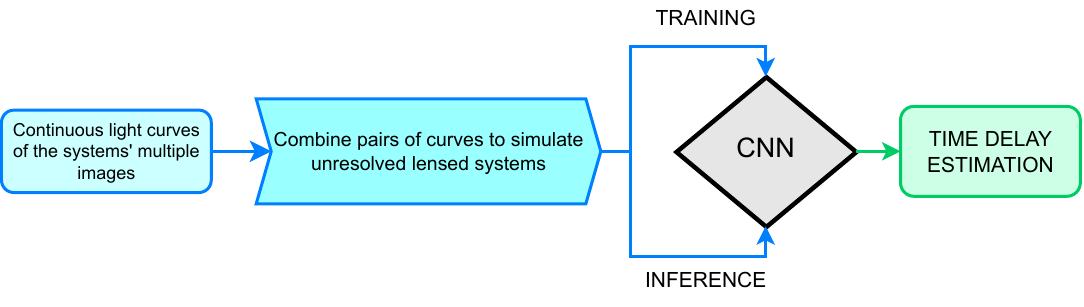}
  \caption{Overview of the proposed methodology.}
  \label{fig:cnn_scheme}
\end{figure}

Our work follows the latter approach: the training data are manually constructed via a physics-based lens simulator described in Section \ref{sec:MC}. The benefit of such an approach is that, depending on the available computational resources, arbitrarily large datasets can be generated. On the other hand, the obvious downside is that the performance of the model at inference time, i.e. when tested on real data, will be strongly dependent on the degree of fidelity of the artificial data with respect to the real ones. This problem is also known as \textit{sim2real gap} \citep{s2r1,s2r2} and it is a very important aspect in many disciplines, including robotics and computer vision. In this paper, we show that, given an as accurate as possible simulator
(in the sense specified above), a fully data-driven CNN is able to
retrieve the time delay from a \textit{single} time series
representing the overlap between two unresolved light curves. The
approach is modular in the sense that, if a more precise simulator is available, it can simply be replaced and used to generate new data to retrain our CNN models with it. 

Furthermore, available models in the literature have yielded state-of-the-art results in the context of time-series classification and regression. Here, we use such
approaches with very little variations compared to their original form; significant changes in the architectures will not be needed in order to obtain the desired results, even in presence of data generated from a different simulator. In the following, we motivate our choice of CNNs as our data-driven models and we introduce the basic principles behind their architecture. Then, we provide detail on the design choices and  our training procedure.

\subsection{Convolutional Neural Networks}

CNNs \citep{cnn1} have been initially proposed in the context of Computer Vision
applications, such as image classification \citep{cnn2} and segmentation \citep{cnn3}. 
They differ from standard fully-connected neural networks (where each node in a certain layer is linked to any other node in the subsequent one) since they implement a convolution operation conferring them two biologically-inspired properties, namely weight sharing and local connectivity. The first results in the same weights being applied repeatedly to different areas of the input data, whereas the second imposes that the action of such weights is realized only locally, on small regions of the input space. Modern CNNs consist of multiple stacked \textit{layers} implementing the aforementioned operation in a hierarchical fashion.

Besides Computer Vision, CNNs have been also fruitfully applied to
time series regression and classification \citep{ResNet,InceptionTime}. The main difference compared to the standard CNNs applied to Computer Vision problems is that, in the case of time series, the filters used by the neural network are now one-dimensional. The choice of such networks for time series analyses is motivated by the structural assumptions (or inductive biases) at the basis of the design of CNN models. Indeed, deep CNNs implement a series of convolutions at each level of the hierarchy along their depth. They work by extracting \textit{local} features from the input raw data, whose representation assumes increasingly higher levels of complexity as we move along the deeper
layers of the network. Our basic assumption is, therefore, that eventual traces of the magnitude of the time delay between two curves
manifest themselves at a local level, motivating the choice of CNN as feature extractor. The application of CNNs to the problem of
time-delay estimation is described in the following paragraph.

\subsection{Time-delay estimation with CNNs}

The input of the CNN models consists of a time series representing an
unresolved quasar light curve $\mathbf{x}=\{x_t\}_{t=1}^T$ where $T$
is the length of the sequence. The output of the model is a real
number $\hat{y}\in\mathbb{R}$ representing the time delay between the
two superimposed curves that went into creating the input time series.

Models are trained by a generated dataset
$\mathcal{D}=\{\mathbf{x}_i,y_i\}_{i=1}^N$, where $N$ is the total
number of training examples and $y_i$ is the ground-truth time delay
associated with the i-th training instance. This initial dataset is
split into three parts, namely a training dataset, $\mathcal{D}_{tr}$,
a validation dataset, $\mathcal{D}_{val}$, and a testing dataset,
$\mathcal{D}_{ts}$. The first is used to train the weights of our
model, the second to check the generalization performance during
training, and the last one to evaluate the model once the training
phase is terminated. Such a splitting is necessary to monitor the
occurrence of the so-called overfitting phenomenon, i.e. when the
neural network simply memorizes the training set and does not
generalize outside the training distribution. 
Figure \ref{fig:cnn_scheme} summarises our methodology. The artificial data produced as will be detailed later are used to train the CNN model. Inference is then performed on the original non-resolved light curves using the trained model. 

The mean-squared error (MSE) is used as a loss function, i.e. to
measure the error the network is making in predicting $\hat{y}$
instead of $y$:
\begin{equation} \mathcal{L}=\frac{1}{N}\sum_{i=1}^N(\hat{y}_i-y_i)^2.
\end{equation}
During training, the weights of the network are varied so that the value of this loss is minimized. This process is realized by the  \textit{back-propagation algorithm}, which allows for the efficient calculation of the gradients of the loss function with respect to the weights in the network. The optimization algorithm used for minimizing the loss is called stochastic gradient
descent, and the popular Adam \citep{adam} variant of this algorithm is
used here:  we use it with a learning rate of $10^{-3}$; a batch size of 50 is selected. We also periodically evaluate the network on the
validation set during training and check its performance in terms of
MSE. As commonly done in practice, whenever the validation loss
decreases, the weights of the network at that step of training are
saved.

\subsection{Models}
Three different CNN models are tested to perform the time-delay estimation task. The first two, namely ResNet \citep{ResNet} and InceptionTime \citep{InceptionTime}, are the results of recent research efforts in the area of time series classification and are adapted to our scope with minimal changes compared to their
original implementation.
The main innovation introduced by these two models stands in their use of the residual connections, which facilitate the propagation of gradients even in relatively deep networks, without incurring in the so-called vanishing/exploding gradient problems, by introducing elements like bottleneck layers and allowing for multiple filters of different lengths to be applied simultaneously to the same input time series. 

Inception modules have a slightly more complex structure compared to Resnet modules which are mainly made of simple fully convolutional layers \citep{McNeely}. For more details on the specifics of the architectures of these two models we refer the interested reader to the works of \cite{ResNet} and \cite{InceptionTime}.
We adapt the original implementations of these models to the regression setting by changing the dimension of the output layer to one. 
To the best of the authors' knowledge, this is the first time ResNet and InceptionTime have been applied to a problem involving time-series in the context of cosmology and we hope our work can be inspirational for future applications of such models.

The third model, simply labelled generically "CNN" in the following, is a relatively standard deep fully convolutional network without residual connections and it is fine-tuned to maximize the performance on the validation set: in this way we can compare the results obtained with a fine-tuned network with those from not-fine-tuned ones. This CNN consists of 13 convolutional layers, each with 14 filters of size 17 and 3 linear layers mapping the output of the convolutional blocks into the final output. Batch-Normalization \citep{ioffe2015batch} and ReLU \citep{agarap2019deep} activations functions are used after each convolutional layer.

\section{Generation of mock light curve datasets}
\label{sec:MC}


We use mock data of the lensed systems to generate the training set
required by our deep learning algorithm. Specifically, the
\textit{MOLET}\footnote{\url{https://github.com/gvernard/molet}} software package \citep{molet} is used to generate light
curves of GLQs multiple images.
\textit{MOLET} allows to include the
microlensing effect that affects the time delay estimates and is thought to be constantly present \citep{Cosmograil2020}. Fig. 1 in \citep{molet} illustrates the flow chart of \textit{MOLET}: it receives in input different information such as cosmological and astrophysical parameters of the lensed system, the intrinsic light curves of the source, magnification maps, the telescope parameters and a realistic observational plan accounting for daily and seasonal gaps.

We chose to simulated two systems with different features, broadly representative of the various known lensed quasars. The first one,
hereafter denoted as system $A$, is a basic test-system for an AGN
point source, with a simple intrinsic variability and with
microlensing. The second system is RXJ 1131-1231 \citep{RXJ}, hereafter denoted
as system $B$ for brevity. It consists of four multiple images: here
they will be combined in pairs to mimic an unresolved doubly-imaged quasar.
%

\begin{figure}
  \centering
  \includegraphics[width=\columnwidth]{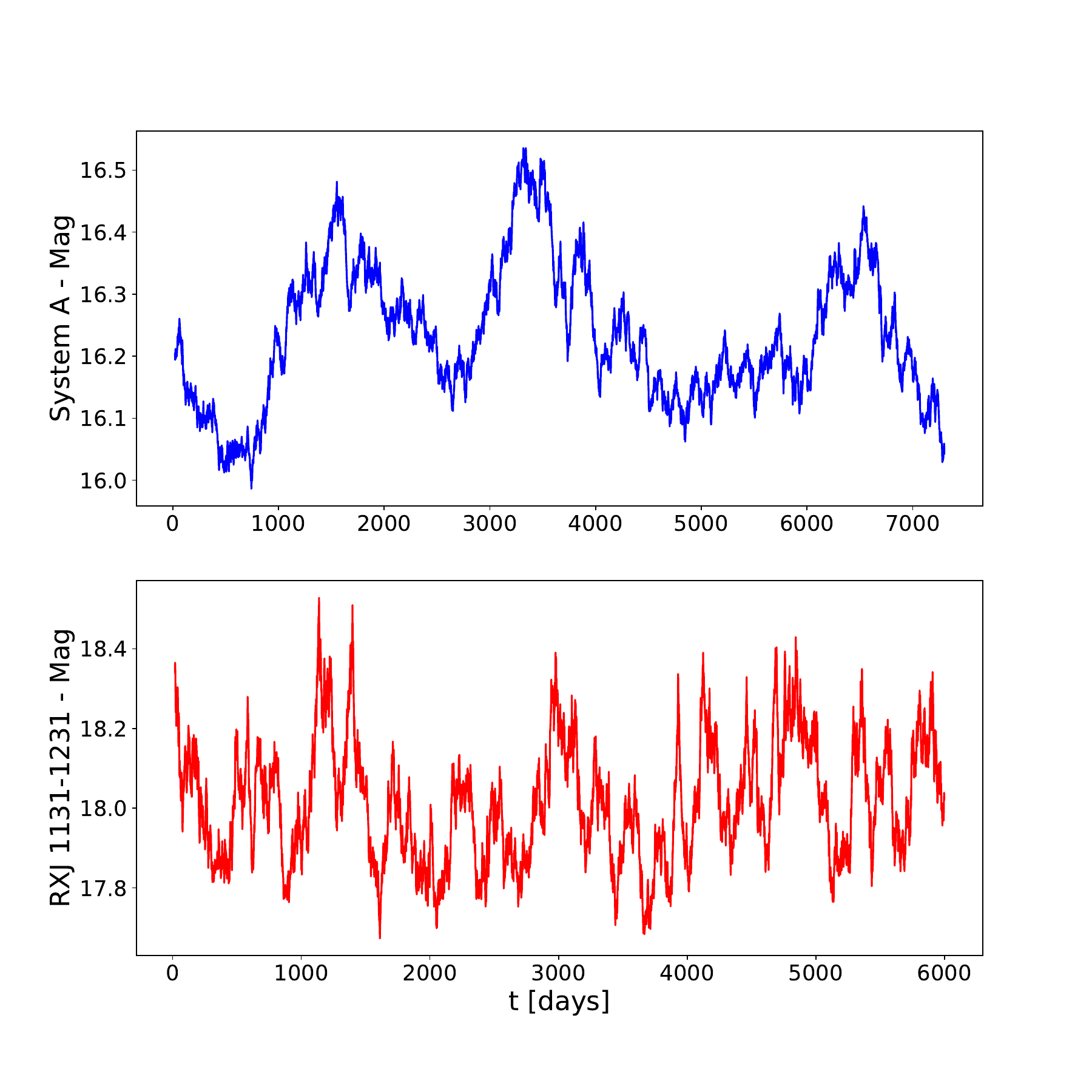}
  \caption{Assumed intrinsic variability of the quasar point source
    of system $A$ (top) and \textit{RXJ 1131-1231} (bottom).}
  \label{fig:intrinsic_curves}
\end{figure}

To properly simulate the observed light curves, \textit{MOLET} needs the intrinsic
variability of the quasar sources. The two systems that we analyze feature
distinct intrinsic variabilities, as shown
in Fig. \ref{fig:intrinsic_curves}, representative of two typical regimes for lensed quasars.  The magnification maps needed by the second step of the \textit{MOLET} simulation are available for both
systems from the GERLUMPH resource \citep{gerlumph}.  Finally, the last
step of the \textit{MOLET} run accounts also for the assumed instrumental gaps (daily or
seasonal) to simulate a realistic campaign from an optical telescope.
More details on the simulation of system $B$ can be found in \cite{molet}.

To build our mock up data, we run \textit{MOLET} several times, by fixing the value of the input time delay and extracting for each run random numbers in a given range of time delays (0-40 days): in this way we obtained 2000 simulations of system \emph{A}, and 8000 simulations of system $B$. Each simulation produced four resolved light curves,
one for each multiple image. The original output of the mock
simulation is continuous light curves: these must then be made discrete and noise must be added to them as well.
Since the goal of our work is to measure the time delay in unresolved
doubly-imaged systems, the separate continuous mock data curves have to be added up in pairs to obtain a single light curve that mimics an unresolved doubly-imaged GLQ. In
this way, a total of six realisations of a single unresolved light
curve of a doubly-imaged quasar are obtained from the individual light
curves of the four images of the original GLQ. Each of these six light
curves is characterised by a different time delay, which is given by
the absolute difference of the time delays of the underlying light
curves.
The combination of the light curves is performed adopting the
following definition of the magnitude of a generic source $X$:
\begin{equation}
  \label{eq:mag-flux-relation} \mathrm{mag}_X = -2.5 \log_{10} F_X + K,
\end{equation}
where $F_X$ is the \emph{flux} of the source $X$, i.e. the energy per
unit time per unit area incident on the detector, and the constant $K$
defines the zero point of the magnitude scale. When two images $A$ and
$B$ of a GLQ cannot be resolved, a single image $O$ will appear on the
detector, with a flux given to a first approximation by the sum
$F_O = F_A + F_B$ of the fluxes of the two overlapping
images. According to Equation \eqref{eq:mag-flux-relation}, the
corresponding magnitude is:
\begin{equation}
  \label{eq:mag-flux-sum} \mathrm{mag}_{O}(t) = -2.5 \log_{10}
  \left( F_A(t) + F_B(t + \Delta t) \right).
\end{equation}
If $B$ is delayed by $\Delta t$ with respect to $A$, the information
about the time-delay $\Delta t$ should still be present in the
features of the light curve of the image sum $O$. Such information needs to be properly extracted as it is hidden by the microlensing effect.

\begin{figure}
  \centering
  \includegraphics[width=\columnwidth]{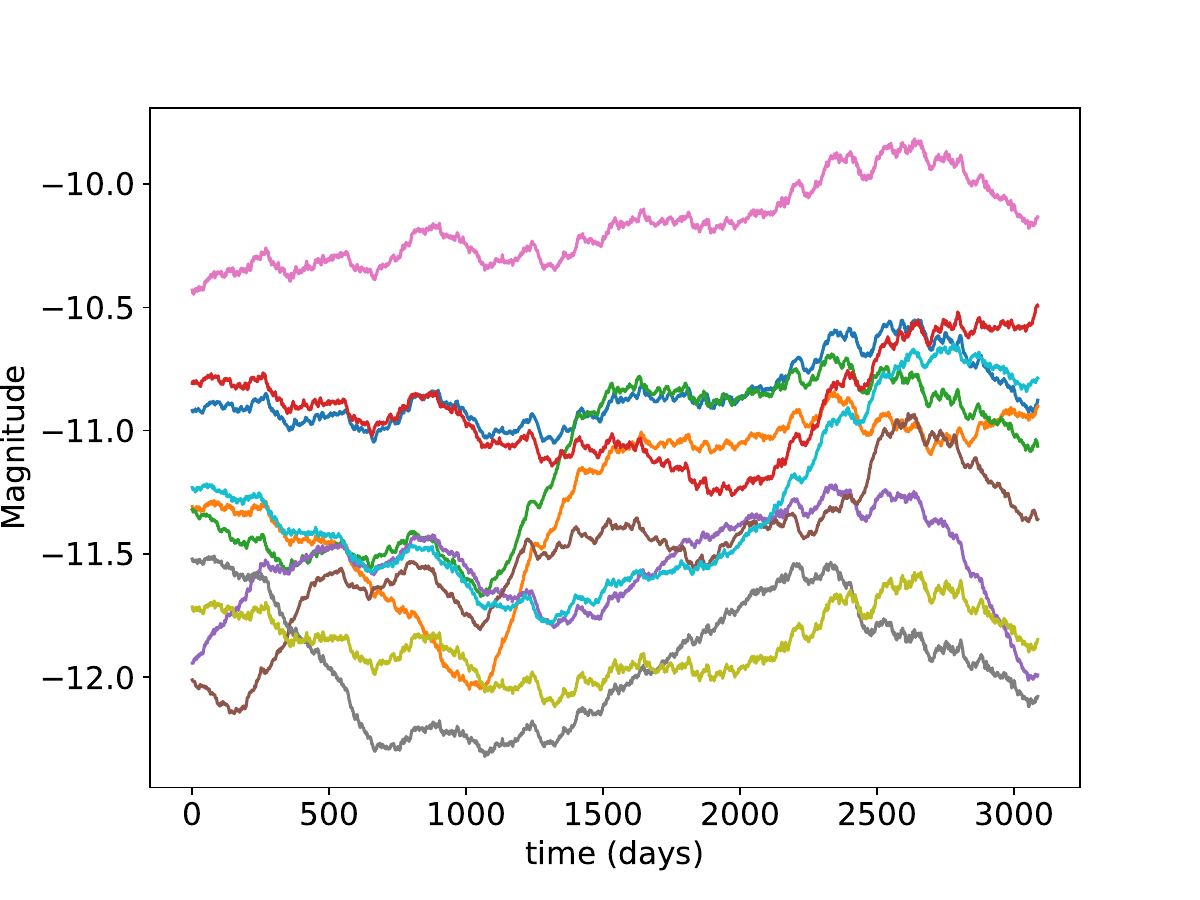}
  \includegraphics[width=\columnwidth]{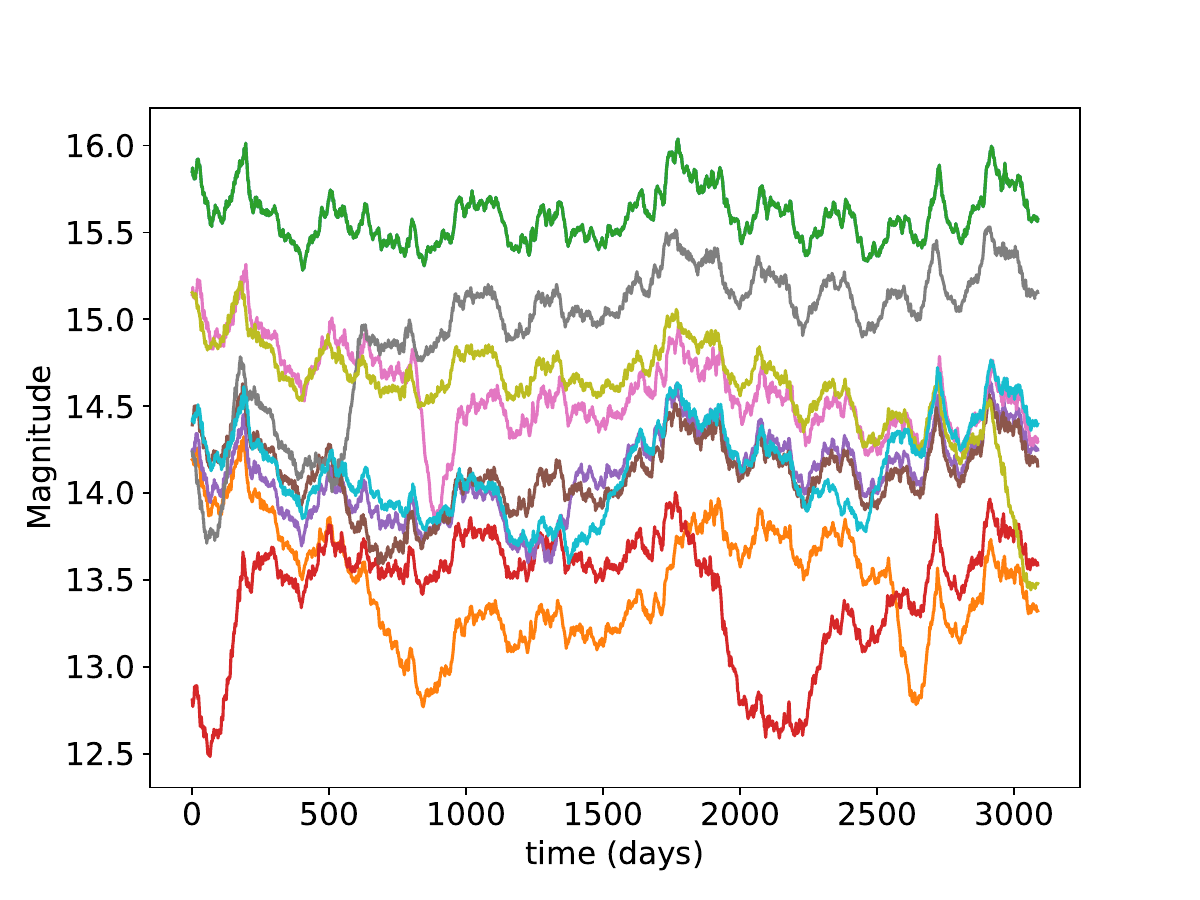}
  \caption{Continuous sample light curves for system $A$ (top) and system $B$ (bottom). The light curves have been randomly chosen from the whole simulation sets for illustrative purposes. Therefore, the observed differences are due to the fact that each of them comes from a different simulation, with different microlensing effects. The simulation period covers about 10 years of data taking.}
  \label{fig:ex12}
\end{figure}

Using all the various possible combinations in pairs of the multiple
images has allowed to get $6\cdot 2000=12000$ unresolved light curves
for system $A$ and $6\cdot 8000=48000$ unresolved light curves for system $B$. Figure \ref{fig:ex12} shows ten light curves for each system, randomly chosen by the whole simulation sets, for illustrative purpose.

\begin{figure}
  \centering
  \includegraphics[width=\columnwidth]{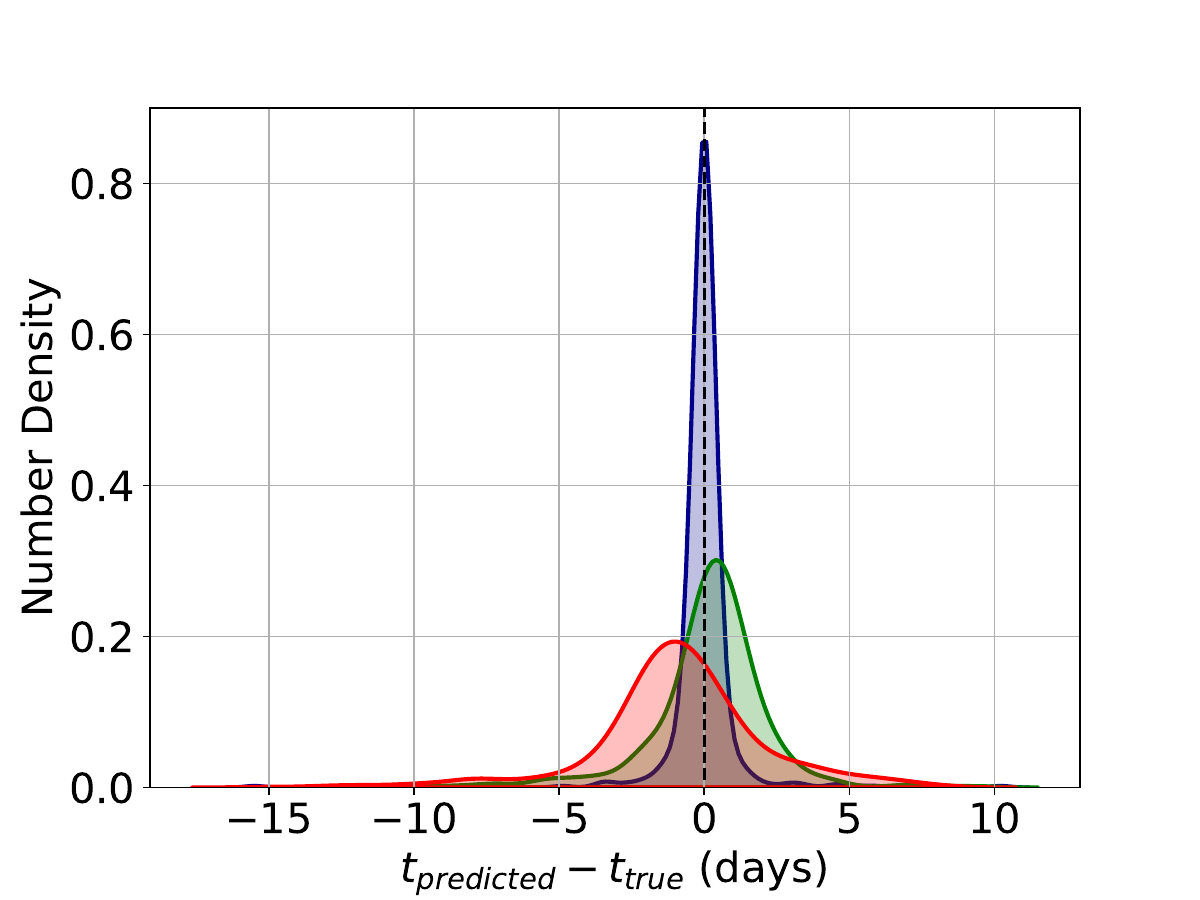}
  \includegraphics[width=\columnwidth]{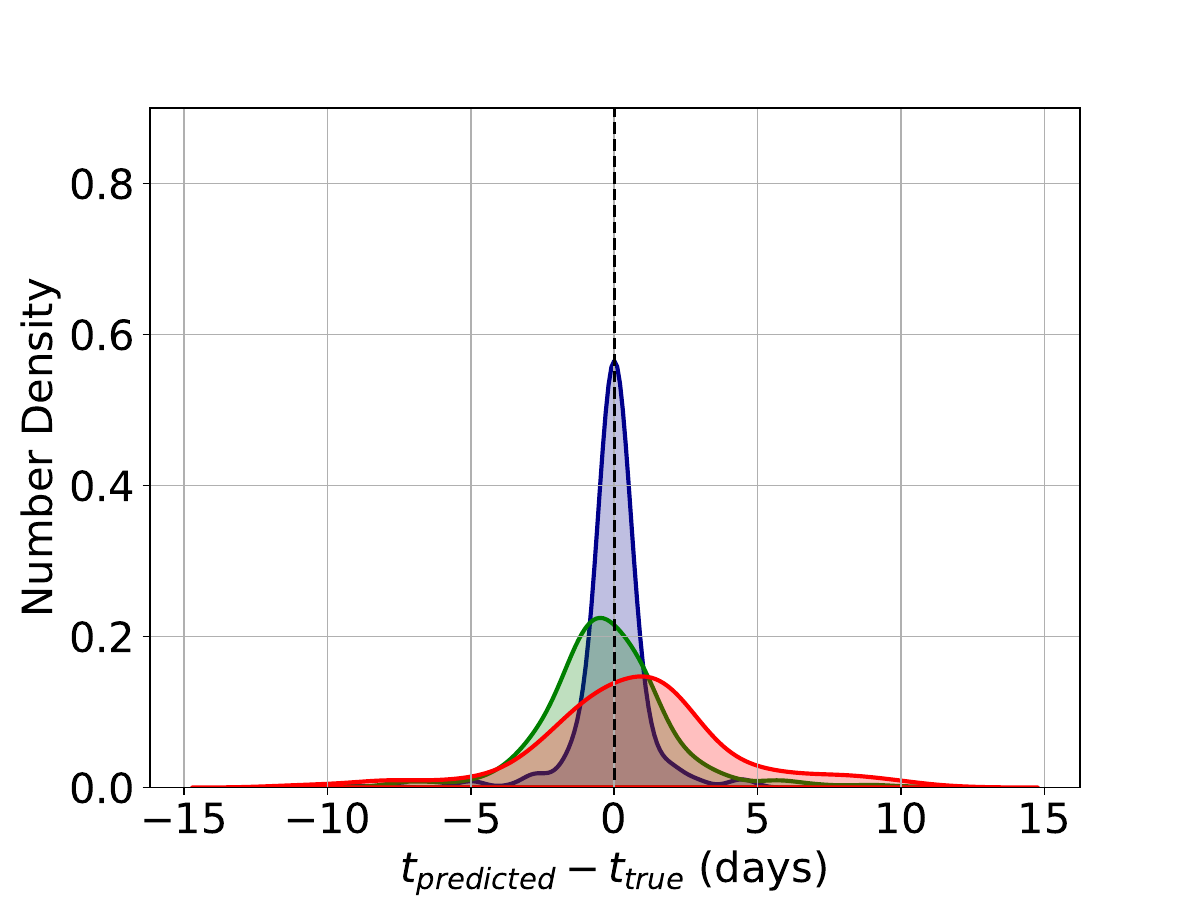}
  \caption{Kernel Density Estimation of the error distributions on
    the test set provided by our CNN model (blue), InceptionTime (red) and
    ResNet (green), for the system $A$ (top) and system B (bottom).}
  \label{fig:hist}
\end{figure}

\begin{figure}
  \centering
  \includegraphics[width=\columnwidth]{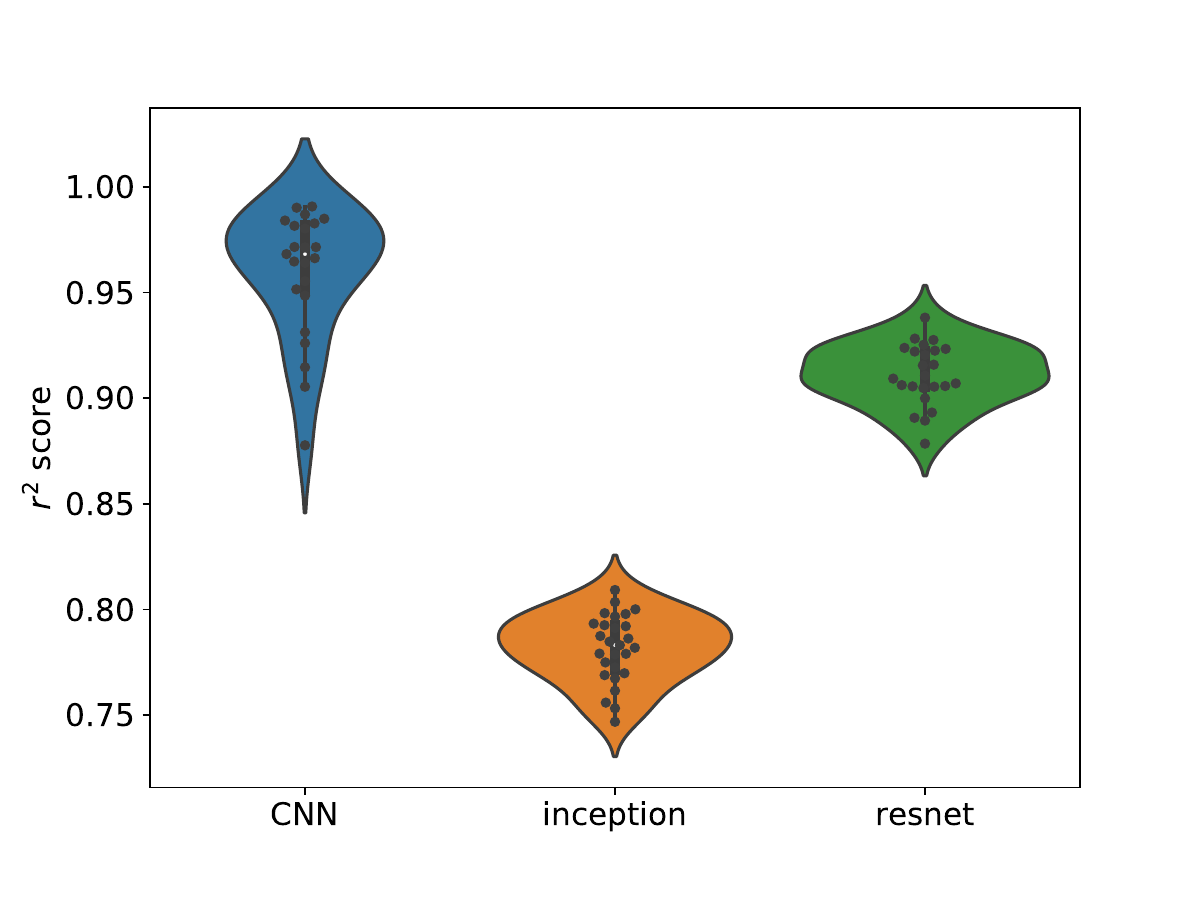}
  \includegraphics[width=\columnwidth]{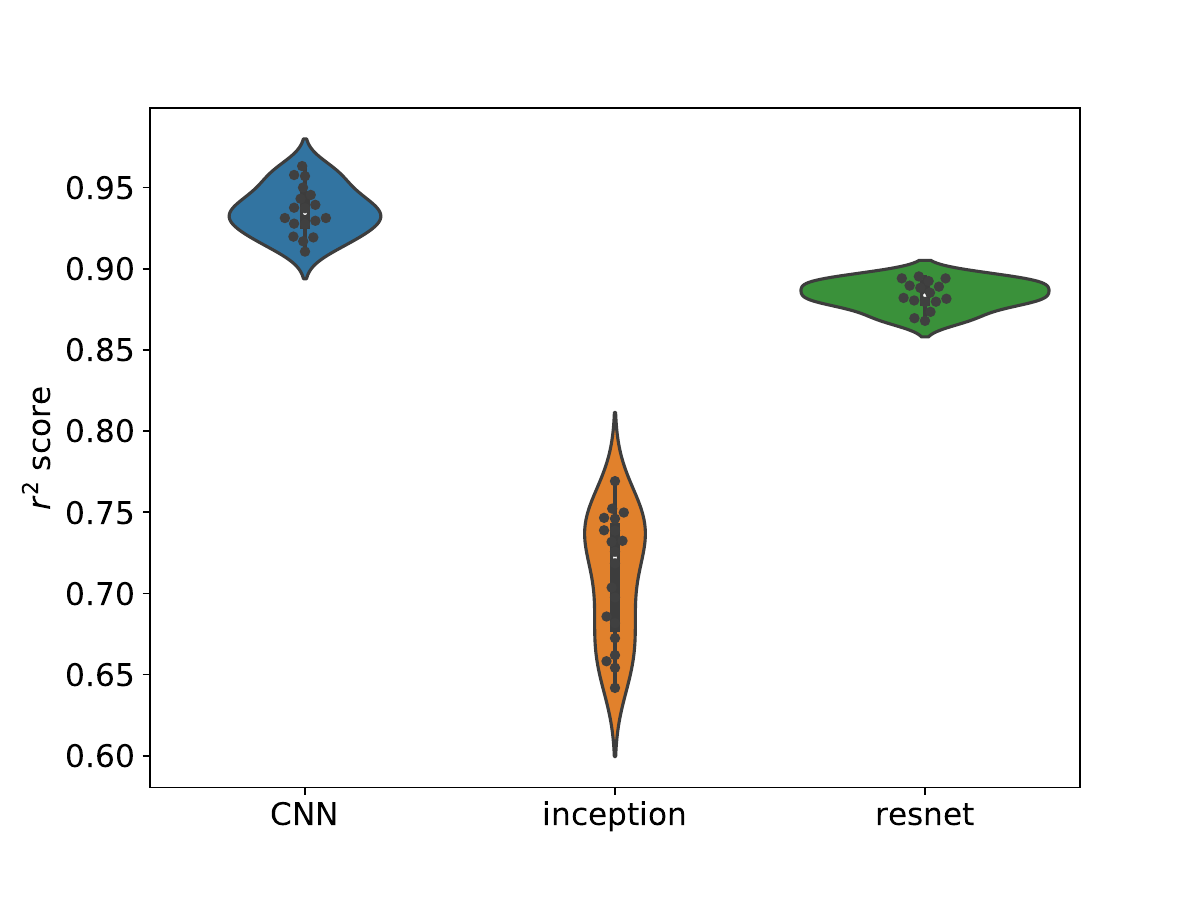}
  \caption{$r^2$ score distributions provided by the three
    considered models, for system $A$ (top) and system $B$ (bottom).  }
  \label{fig:violin_plot_r2}
\end{figure}

\section{Results}
\label{sec:test_approach}
Here we report our results on the performance of the three proposed CNN architectures on the \textit{test data}, i.e. light curves that our models have never seen during their training phase. The goal is to verify their level of generalization on \textit{new} test curves extracted from the same distribution as the training data. Note that, since the test data are generated with the same simulation engine used to produce the training set, here the out-of-distribution generalization ability of our
algorithm is not tested, i.e. we do not test its robustness with respect to the sim2real gap phenomenon. 

The performance of the models on both systems $A$ and $B$ is 
analysed at different levels. The first evaluation studies the error distribution $t_{predicted}$-$t_{true}$ between the predicted time delay and the ground truth, i.e. the time delay used to construct the training set. Given that the optimizer we use  has a stochastic component in its operation, the outcome of the training can be slightly different from one run to another, so several training runs are performed for each model and, for this experiment, the best performing one for each architecture is selected. The kernel density estimation
\citep{Silverman, Scott} is used to approximate the error distributions
for each model and each system. The results are shown in
Fig. \ref{fig:hist}. For both systems, the distribution provided by the CNN
model is much narrower and symmetric around zero, that is the predicted time delay is closer to the input time delay. ResNet seems to
outperform InceptionTime, even though both provide broader and more
skewed curves.

As a second evaluation, the distribution of the $r^2$ coefficient of
determination \citep{Silverman, Scott} is investigated to establish to
what extent prediction and ground truth are aligned. The $r^2$ score
represents the fraction of the variance of a dependent variable that
can be predicted from the independent variables and is then a metric
used to evaluate the goodness of fit of a regression model, a value of
1 being the ideal case. For each system, each model is trained 20
times and the performance of each trained model is assessed on the test
set in terms of the $r^2$ score. Figure \ref{fig:violin_plot_r2}
shows the resulting probability density of the data at different
values, the so-called "violin plots": in both cases, our CNN yields
$r^2$ scores closer to 1. However, in the case of system $A$, the
$r^2$ distribution of our CNN is characterized by a slightly larger
variance, resulting in an overlap with the ResNet $r^2$
distribution. InceptionTime is outperformed by the other two baselines
in both systems and, in the case of system $B$, it produces a high
variance distribution with realizations ranging between a minimum of
0.6 and a maximum of 0.8 $r^2$ scores.

In summary, the analysis in terms of error distribution and $r^2$
score highlights that all models provide very good performance on both
systems. It is important to emphasize that InceptionTime and ResNet
were not fine-tuned, in order to keep them the same as the original
implementations from the literature. This was done on purpose to
showcase the flexibility of these models to work on very heterogeneous
types of data. We therefore expect their performance to further
improve with more careful architectural and hyper-parameter design
choices.

\begin{figure}
  \centering
  \includegraphics[width=\columnwidth]{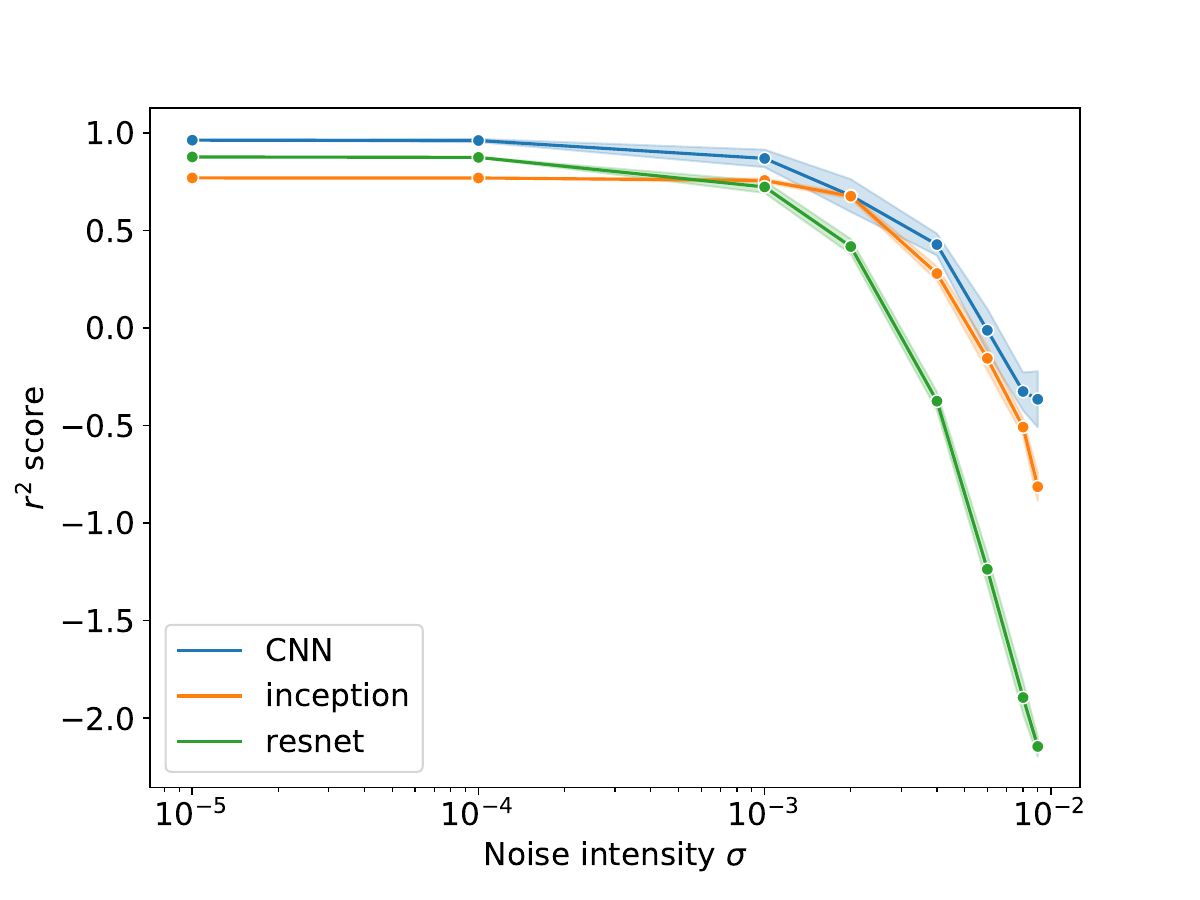}
  \includegraphics[width=\columnwidth]{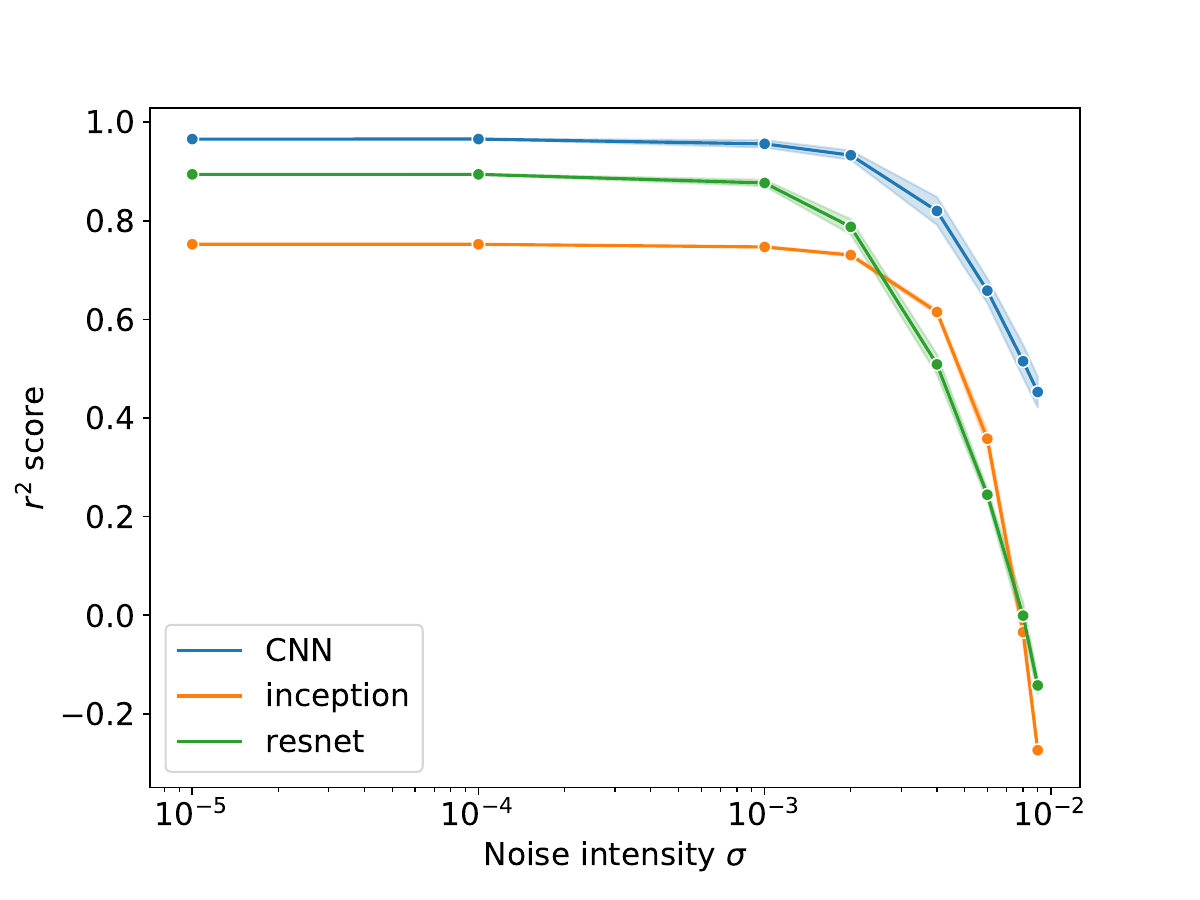}
  \caption{$r^2$ scores for the three models as injected noise
    standard deviation increases, for system $A$ (top) and system $B$ (bottom).}
\label{fig:r2_noise}
\end{figure}

After having verified that the proposed methods generalize well on the test set, their robustness is assessed when noise is injected in the data. To this extent, we perturb the original time series with a zero-mean Gaussian noise with standard deviation $\sigma \in \{0.00001,0.0001,0.001,0.002,0.004,0.006,0.008,0.009\}$. These values have been selected so that the main structural properties of the resulting light curves are not altered too much by the injection of noise and their macroscopic visual appearance is roughly preserved. This operation effectively introduces a bias between training
and testing distributions, whose severity depends on the intensity of
the perturbation. Figure \ref{fig:r2_noise} shows how the $r^2$ score of each model
decreases as the standard deviation of the gaussian noise
increases. Again, generally the CNN outperforms the other
models. However, in system $A$, its curve tends to align with the
InceptionTime one as the noise level increases. Interestingly, the CNN
model seems to be more robust on system $B$, providing satisfactory
values of the $r^2$ score even for relatively large noise standard
deviations. In general, the performance of the three models on system A seems to be much more sensitive to noise perturbation than that obtained by the models on system B. We do not have a satisfactory explanation for that yet: one likely possibility is that it depends on the difference in the size of the datasets associated with the two systems, but more investigations are needed to draw a firm conclusion.

\begin{figure}
  \centering
  \includegraphics[width=\columnwidth]{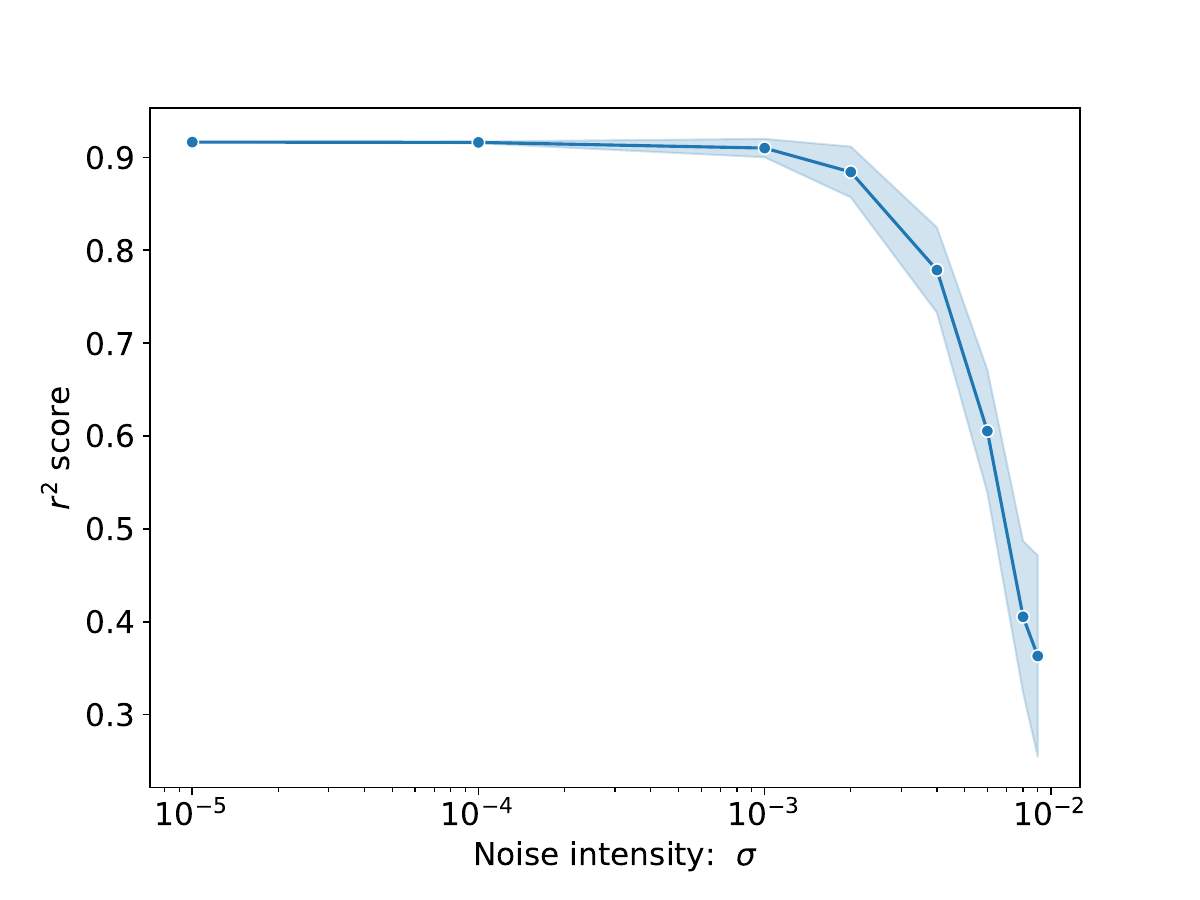}
  \includegraphics[width=\columnwidth]{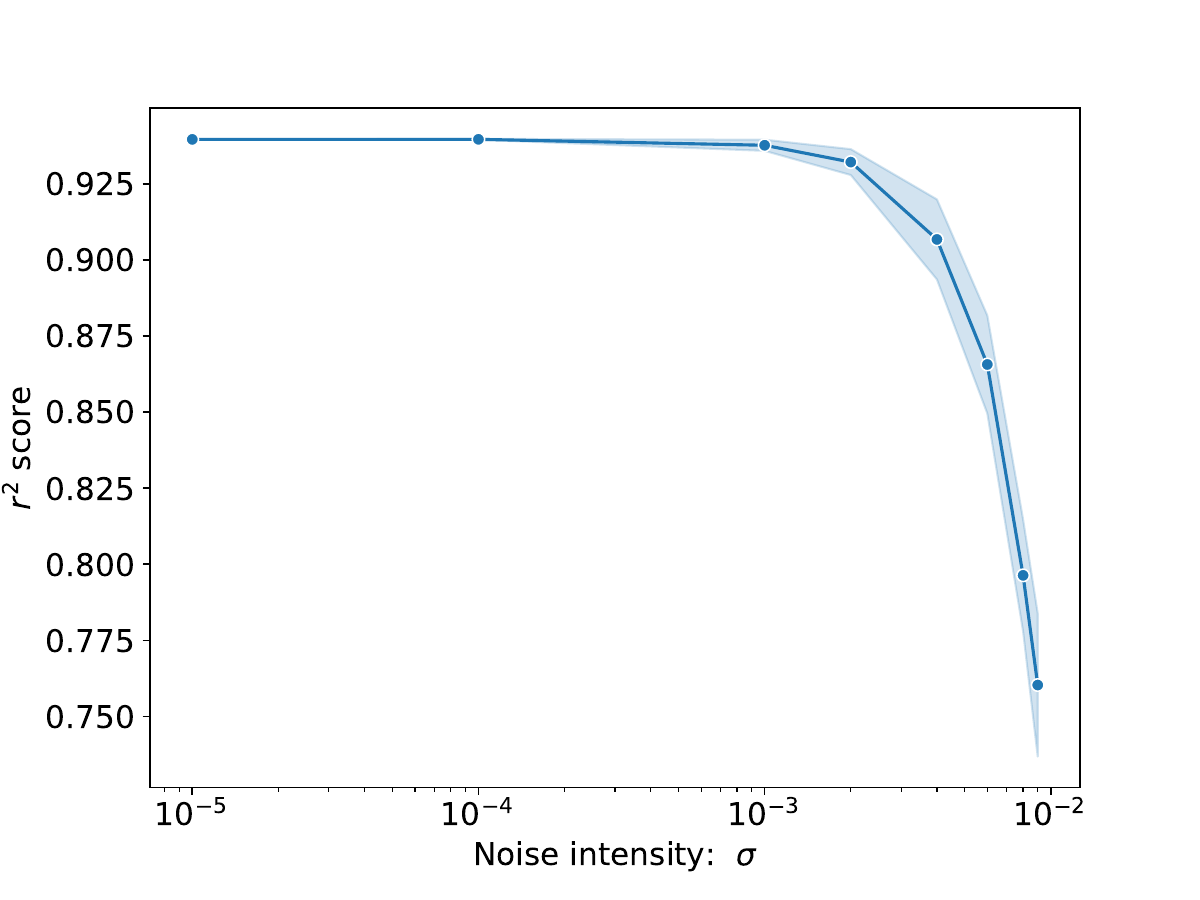}
  \caption{$r^2$ scores for our CNN model trained with random noise
    injection as noise standard deviation increases, for system $A$ (top) and system $B$ (bottom).}
  \label{fig:r2_noise_CNN}
\end{figure}

As a last experiment, we study how the previous analysis changes if we
train our best performing CNN on data perturbed by noise at variable
standard deviations. To do so, at training time for each batch of
data, we randomly sample a value for the noise standard deviation from
a uniform distribution between $10^{-5}$ and $10^{-2}$ and we feed the
corrupted data into the network. Once training is terminated, we
repeat the previous experiments. The results for system $A$ and system
$B$ are shown in Fig. \ref{fig:r2_noise_CNN}.

The result of injecting noise at training time is a model which is
more robust to perturbations in the test data. On the other hand, this
positive effect is obtained at the price of a slight degradation in
performance when the noise level is low. This experiment suggests that
randomly injecting noise in the data at training time in this case
represents an effective strategy to obtain more robust models.

In light of the presented experiments, the proposed CNN architectures
appear to guarantee satisfactory performance on the task of
predicting the time delay from unresolved quasar light
curves. Overall, the CNN model proposed here seems to outperform the
others even though more fine-tuning and more carefully designed
choices might eventually close this gap.

\section{Conclusions}

In this work, a new class of deep learning-based methods has been used to extract the time delay between GLQs unresolved light
curves. The method is motivated by the existing tension on the estimated values of $H_0$, which can only be resolved by reducing the uncertainty on $H_0$. In this respect, there is the necessity to increase the number of analysed GLQs by processing and extracting information from unresolved quasar light curves, which can be typically provided by small/medium telescopes. The obtained results show that the proposed approach performs nicely on mock data describing two different lensed quasar systems. Moreover, it has several advantages compared to classical approaches: first, it is designed to process unresolved light curves which represent the majority of the data that small/medium sized telescopes are expected to provide.
%
Second, the method is fully data-driven: its
performances scale easily with the dataset size and it makes
little to no assumptions on the nature of the time-delay estimation
problem.

On the other hand, the current implementation is still affected by
some limitations that open new interesting research questions. Most
importantly, the method is highly reliant on the quality of the
simulated data and on their level of fidelity with respect to real
quasar light curves. It follows that, when applying the proposed approach to real data, a degradation in performance shows up. This
problem is a manifestation of the sim2real gap phenomenon described in
Sec. \ref{sec:test_approach}.  
In order to alleviate it, the gap between simulated and real data must be reduced as much as possible. One popular strategy to cope with this problem is the technique of \textit{domain randomization} \citep{tobin2017domain,2018,chebotar2019closing,prakash2020structured}, where large and very diverse datasets are generated by randomizing the parameters of the simulator with the hope that, when the model is deployed on real data, the new observations will somewhat close to the randomized simulations the model has been trained on.

On the other hand, the sim2real gap phenomenon, and in particular the size of the discrepancies, can be exploited to assess
the quality of the simulated data: the better the performance of the
network, the closer the simulator grasps the details of the data it is
trying to emulate. 

Lastly, as a further improvement to the method that will be investigated in future works, is to enable it to process also irregularly sampled time series which are indeed commonly encountered in the context of cosmological and astrophysical applications, because of unavailabilities or outages of the instruments, for example. 
Finally, a future extension of this work will deal with systems having $N>2$ images: in fact, a large amount of quadruply-imaged GLQs is expected to be detected in the future \citep{ShuY}. We aim to investigate these new research directions in future works.

\section*{Acknowledgements}
Part of this work was supported by the German
\emph{Deut\-sche For\-schungs\-ge\-mein\-schaft, DFG\/} project number
\texttt{Ts~17/2--1}.  GV has received funding from the European
Union’s Horizon 2020 research and innovation programme under the Marie
Sklodovska-Curie grant agreement \texttt{No 897124}. Authors
acknowledge support from the ``Departments of Excellence 2018-2022''
Grant (\texttt{L.~232/2016}) awarded by the Italian Ministry of
University and Research (\textsc{mur}).

\section*{Data Availability}
The data underlying this article will be shared on reasonable request to the corresponding author.



\bibliographystyle{mnras}
\bibliography{biblio} 

\bsp    
\label{lastpage}
\end{document}